# Toward Microgravity Mass Gauging with Electrical Capacitance Volume Sensing: Sensor Design and Experiment

# Toward Microgravity Mass Gauging with Electrical Capacitance Volume Sensing: Sensor Design and Experiment

Matthew A. Charleston[1*], Shah M. Chowdhury[1], *Member, IEEE*, Qussai M. Marashdeh[1], *Senior Member, IEEE*, Benjamin J. Straiton[1], Fernando L. Teixeira[2], *Fellow, IEEE*

*Abstract*— The use of capacitance sensors for fuel mass gauging has been in consideration since the early days of manned space flight. However, certain difficulties arise when considering tanks in microgravity environments. Surface tension effects lead to fluid wetting of the interior surface of the tank, leaving large interior voids, while thrust/settling effects can lead to dispersed two-phase mixtures. With the exception of capacitance-based sensing, few sensing technologies are well suited for measuring annular, stratified, and dispersed fluid configurations. Two modalities of capacitance measurement are compared – Electrical Capacitance Volume Tomography (ECVT) and Electrical Capacitance Volume Sensing (ECVS). ECVT is a non-invasive imaging modality first introduced in 2006. ECVS is a measurement modality introduced in this paper that is derived from ECVT technology but does not reconstruct an image as part of the mass measurement. To optimize the design of future capacitance-based spherical tank mass gauging sensors, different electrode plate layouts are evaluated in a mass gauging context. Prototype sensors are constructed, and experiments are conducted with fluid fills in various orientations. The plate layouts and their effect on the performance of the sensor as a fuel mass gauge are analyzed through the use of imaging and averaging techniques.

*Index Terms*—Capacitance, Electrical Capacitance Volume Tomography, Microgravity Mass Gauging, Propellant, Sensor

## I. Introduction & Background

PROPELLANT mass gauging has long been a critical measurement in rocketry. Although a simple level measurement problem on the ground, mass gauging in microgravity is a complex challenge. In a low-gravity environment, surface tension effects distort the liquid/gas phase interface, and momentum effects can create dispersed fluid configurations [1]. Liquid level sensing modalities like ullage pressure correlations, ultrasonic interface detection, or coaxial capacitance probes have been used in a number of flights but require wasteful settling thrusts to generate an accurate measurement [2]. Other established methods such as propellant mass flow integration, or bookkeeping, are only applicable when thrusting, accumulate error over time, and are unable to detect propellant leaks [1]. The pressure-volume-temperature (PVT) method, based on adding a pressurizing gas and using the ideal gas law, can gauge propellant mass but loses accuracy as the amount of fluid in the tank decreases [3]. Thermal propellant gauging, on the other hand, increases in accuracy at low fill levels but makes assumptions on fluid position [3]. Radio frequency and acoustic modal gauging methods are under development but require detailed understanding of the fluid configuration [4] [5]. Despite decades of research, microgravity propellant mass gauging remains an unsolved problem and is an active area of focus for NASA [6] [7]. A new, on-demand, passive measurement technique that is accurate irrespective of fluid configuration is clearly needed for future orbital and interplanetary missions [8].

Capacitance-based techniques offer a promising method for mass gauging that eliminates some of the problems with conventional techniques. In capacitance sensors, an electrode or plate is individually excited while an electric current from every other plate is measured. The measurement process is repeated sequentially until all combinations of plates are measured. Using this strategy, the number of independent measurements, or "sensing channels", M, is given in terms of the number of plates, N, as $M = N(N-1)/2$ [9] [10].

Because it contains many independent measurements, data collected using this method contains information about the position of the fluid as well as the quantity, making it a more robust technique for dynamic, unsettled, or sloshing fluid configurations, and allowing it to measure the fluid volume without a priori assumptions on how the fluid will settle in microgravity conditions.

Capacitance-based fuel mass gauging has seen substantial scientific interest. In the 1960s, Capacitance level sensing probes were implemented on the Saturn V and a whole tank capacitance gauge concept was introduced [1] using what has been referred to as a "spatial regularization" approach [11]. However, in [12] it was proven that the parallel field approach of spatial regularization techniques is merely a superposition of conventional electrical capacitance tomography techniques and does not offer any advantages.

Currently, two principle methods exist to reconstruct capacitance data into a mass fraction for microgravity: either the capacitance data is converted to an image through a technique known as Electrical Capacitance Volume Tomography (ECVT) and then the mass fraction is derived from the image, or the mass fraction is directly derived from the capacitance data through an algorithm such as weighted averaging or machine learning [13]. In the second method, no image reconstruction is involved. However, the same volumetric capacitance information is used. Thus, the term

Electrical Capacitance Volume Sensing (ECVS) is introduced. Derived from ECVT, ECVS is a capacitance-based sensing methodology that utilizes the 3D volumetric sensitivity of the electric field to measure the volume fraction or mass fraction of phases inside the sensing region without the use of image reconstruction. While ECVT can provide detailed information on the phase distribution in a tank, the use of tomographic reconstruction techniques to create an image is not necessarily the most accurate method to measure the global fluid mass due to the incorporation of errors through the ill-posed nature of the image reconstruction problem [14]. Additionally, image reconstruction can be a large computational burden. As such, in this paper different algorithms using the same ECV sensor are explored to assess the effectiveness of the sensor design under various techniques.

Several teams have previously investigated the accuracy of capacitance mass gauging sensors in microgravity [15] [16] [17] [13] and sloshing conditions [18] with promising results. These studies, however, use different sensor designs and did not systematically analyze or improve the design to optimize the sensor performance for mass gauging. In this paper we discuss the design constraints and optimizations that can improve the performance of an ECV sensor. Then we construct two types of sensors, conduct mass gauging experiments, and conclude that a dodecahedron style sensor design offers significant advantages over the state-of-the-art.

## II. Sensor Design

A robust and accurate capacitance mass gauge must measure fluid mass independent of fluid position and distribution. However, because capacitance sensing is a soft field technique, the sensitivity of an electrode pair varies in space. The number and specific layout of the electrodes in 3D determines the sensitivity distribution of the ECV sensor. Although post-processing algorithms can compensate for uneven sensitivity, optimizing the electrode geometry is key to the success of a microgravity mass gauge, as the raw capacitance reading should change as little as possible when fluid position or shape changes for a constant fluid mass.

In traditional cylindrical ECT and ECVT sensors, the electrodes are typically a tessellation of rectangles. Other tessellations of polygons such as triangles have been tested [19]. Electrode geometry consisting of tessellations of regular polygons is ideal in ECV sensors because it results in a relatively axially/radially equal sensitivity strength distribution over the RoI, improving image reconstruction and regularization [19].

Mass gauging applications are often for spherical tanks. When imaging a spherical region, new electrode geometry must be considered, as the planar tessellations commonly used in ECT or ECVT applications do not have the same advantages when projected in the spherical domain. Gut [16] tested a design with 24 electrodes consisting of 3 rows of 8 rectangular plates arrayed around a cylinder and then projected onto a sphere. This projected pattern has irregular polygons, and the uneven sensitivity distribution resulted in errors in the calculated volume fraction that varied significantly with rotation of the tank. In order to smooth the sensitivity distribution, geometry that tessellates onto a sphere must be used and regular tessellations are preferred. Regular tessellations utilize the whole surface area with the most rotational symmetry, resulting in a relatively even sensitivity distribution. Examples of this are seen in Fig 1. There are a finite number of regular polyhedra that can be tessellated onto a sphere. These are known from antiquity and referred to as platonic solids [20].

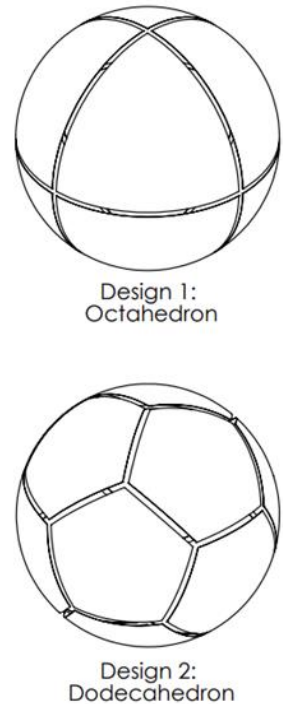

Fig 1. Examples of Spherical Projections of Platonic Solids

TABLE I
RELEVANT PROPERTIES OF SPHERICAL PROJECTIONS OF PLATONIC SOLIDS

|  | Plates | ECVT Channels | Non-Adjacent Channels | Rotational Order |
| --- | --- | --- | --- | --- |
| Tetrahedron | 4 | 6 | 0 | 12 |
| Cube | 6 | 15 | 3 | 24 |
| Octahedron | 8 | 28 | 4 | 24 |
| Dodecahedron | 12 | 66 | 36 | 60 |
| Icosahedron | 20 | 190 | 100 | 60 |

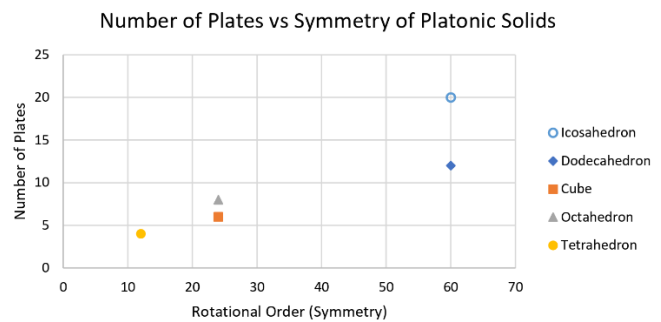

Fig. 2. Number of Plates vs Rotational Symmetry for Platonic Solids

Selecting a projection of a platonic solid ensures a relatively even sensitivity distribution and the efficient utilization of the entire surface area of the sphere. What remains is selecting the specific platonic solid to optimize the properties for the specific application. The relevant properties for each platonic solid are laid out in Table I where rotational order is used to represent the 3D symmetry of the design. Fig. 2 elucidates the relationship between rotational symmetry and the number of electrode plates.

A tetrahedron would make a bad ECV sensor as it has only adjacent plates, and therefore would have poor sensitivity in the center. A cube may make a slightly better ECV sensor as it has three non-adjacent channels and is more rotationally symmetric. The octahedron has four non-adjacent channels and additionally four semi-adjacent channels that only share a vertex and not an entire edge. This, along with the large plate area, makes it a good candidate for study. The dodecahedron is highly rotationally symmetric and has a larger number of non-adjacent channels. Finally, the icosahedron has a dramatically higher number of channels. It is also highly rotationally symmetric and has many non-adjacent plates. However, the large number of plates complicates the problem without increasing the rotational order. Therefore, while the other platonic solids are worth further investigation, this study focuses on the octahedron and the dodecahedron designs for comparison. The octahedron sensor has been studied previously [21] and the dodecahedron is the next choice as the number of plates increases. The ideal sensor should maximize SSNR and minimize spatial variation, having only small signal deviations as fluid position changes for a constant fill level. Although further processing through imaging, sensitivity matrices, and machine learning is possible to increase the accuracy of the mass measurement, the design should first be optimized to output a signal with a linear and stable relationship to fluid mass.

When designing an ECV mass gauge, the principal considerations are the dynamic range, sensitivity distribution, and signal-to-noise ratio. The sensor performance can be influenced by changing the electrode plate areas, number of electrodes, spacing of electrodes, the geometry of the electrodes, and grounding planes [22] [23]. Optimizing these features to improve one performance parameter, however, typically results in a reduction of other performance parameters in response, making ECV sensor optimization a complex task. The Dynamic Range (DR) of the sensor is the measurable difference in capacitance, C, between the 2 phases present in the region, shown in (1) where $n$ is each channel. For a typical 2 phase liquid/gas measurement, the DR is evaluated as the signal difference between a sensor filled with gas and filled with liquid, denoted as empty and full respectively.

$$DR_n = C_{n,full} - C_{n,Empty} \qquad (1)$$

An ECV sensor typically has several different types of channels with different geometric relationships. Because the electric field between a set of two plates is governed by their geometric relationship, different channel types have different sensitivities. Some channel types are sensitive to a large part of the volume and some sensitive to only a small fraction of the volume. This is referred to as the sensitivity distribution. The range of sensitivity can vary as well, with some channels having a very high sensitivity in a small space and a comparatively low sensitivity over a large space, referred to as the sensitivity uniformity [24]. The ideal sensor has a wide sensitivity distribution with high uniformity. Additionally, the placement of objects in certain regions can cause the measured capacitance to decrease due to negative sensitivity [9]. In order to quantify the performance of a sensor with many channels and several channel types, a new parameter is introduced, the Spatial Sensitivity Quotient ($\sigma$), that balances sensitivity distribution with the sensitivity uniformity. The spatial sensitivity is calculated by simulating the sensor design in software and generating a sensitivity matrix $S$ (2) that maps the electric field response $\vec{\zeta}$ of a channel to the presence of fluid in each voxel volume $\delta v_i$ [25]. Each row $n$ of $S$ corresponds to a channel of transmit and receive plates and each column $i$ a voxel in space. $S$ is then used in (3) to calculate the spatial sensitivity quotient of each channel in a volume, where V is the total number of voxels.

$$S_n(i) = \iiint_{\delta v_i} (\vec{\zeta}_{n,transmit} \cdot \vec{\zeta}_{n,receive}) \, dv \qquad (2)$$

$$\sigma_n = \frac{\frac{\sum_{i=1}^{V}|S_n(i)|}{V}}{\max_i[S_n(i)] - \min_i[S_n(i)]} \qquad (3)$$

Equation (3) is intended to weigh both the sensitivity distribution and sensitivity uniformity of a channel. The numerator of the equation reflects the sensitivity distribution by taking the average of the absolute value of the sensitivity of the channel over the measurement region as a whole. The absolute value is taken because even a negative sensitivity region contains information that can be used to quantify an object's position and volume. The denominator of the equation reflects the sensitivity uniformity by calculating the range of sensitivity values in the region. Channels that have plates with one edge adjacent to each other have a very high capacitance and a very high sensitivity. However, the region they are sensitive to is very small and consists mainly of the area around the adjacent plate edges. Objects outside this region can have a strong negative response as well. This wide range results in a singularity, a sharp local change in the sensitivity that causes a large change in measured capacitance as a mass moves only a short distance. Channels with plates that are located opposite each other, across the region of interest, have a lower capacitance but are more equally sensitive to a larger area. These characteristics of adjacent and opposite plates are well studied [23] [26] [9] and measured [27]. Using (3), the high sensitivity and low uniformity of the adjacent plates is balanced with the lower sensitivity and higher uniformity of the opposite plates to better evaluate the effectiveness of a given channel for mass gauging. This metric can then be used to judge the effectiveness of specific sensor geometries. The objective is to maximize the spatial sensitivity quotient for each channel and the sensor as a whole.

An example of this is given in Fig. 3. A sensor consisting of 12 plates wrapped around a cylindrical domain is simulated. An adjacent and opposite plate pair are plotted and the SS for the adjacent channel is 0.0071 whereas the opposite channel is 0.0357 even though the adjacent channel's maximum sensitivity is orders of magnitude higher.

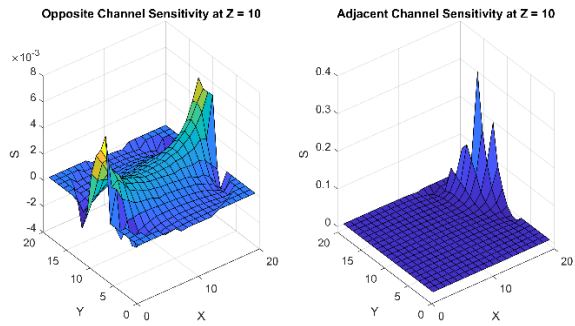

Fig. 3. Sensitivity of Characteristic Channel Types of a Single Plane, Twelve Electrode Capacitance Sensor

In ECVT/ECT, noise in the capacitance signal is generally a property of the measurement system due to the very small changes in capacitance under investigation. It is affected mainly by the amount and type of shielding around the sensor and cables as well as the electronic circuit design. The signal to noise ratio (SNR) is traditionally calculated according to (4), where STD refers to the standard deviation. However, for the same reasons that a spatial sensitivity metric is required, the traditional measurement of SNR is ill suited for use in an ECV sensor. Channels with adjacent plates have a high standing capacitance, and due to that, have dramatically higher SNR than other types of channels. However, this high SNR is not particularly useful from a mass gauging perspective, as the adjacent channels tend to have a lower spatial sensitivity quotient. Furthermore, by only accounting for the full signal in the SNR calculation, the DR of the channel is not factored in. Thus, a channel with high capacitance and low noise would be considered favorable under the traditional definition, even if the change in measurable capacitance from full to empty is quite small. To better judge the performance of the channels and factor in the contribution of environmental noise into the signal, a new equation is introduced, (5) the spatial sensitivity to noise ratio.

$$SNR_n = \left(\frac{C_{n,full}}{STD(C_{n,full})}\right) \quad (4)$$

$$SSNR_n = \left(\frac{DR_n * \sigma_n}{STD(C_{n,full})}\right) \quad (5)$$

When comparing overall sensor performance in mass gauging, the calculated values for each channel are averaged over the total number of channels to make the SSNR applicable to the performance of the sensor as a whole.

A high sensitivity uniformity is important for signal stability under changing fluid positions and has a strong impact on the accuracy of the overall sensor. Poor sensitivity uniformity leads to effects such as the measured volume fraction for a constant volume of liquid changing as the sensor rotates in a gravity field. This effect is combated by increasing the number of electrode plates, allowing the incorporation of many signals together when calculating volume fraction. This increases the overall number of singularities but evenly disperses them around the sensing region to smooth the effect of individual channels on the overall measurement and homogenizes the sensitivity distribution

## III. CONSTRUCTION

The two designs selected for study are displayed in Fig. 1. For both designs, a flanged acrylic hemisphere with an internal diameter of 9.5 inches and a thickness of ¼ inch was used as the tank. The outside of the tank was sprayed with a conductive nickel paint to act as a ground shield, and then coated with a scratch resistant paint for protection. A silicone gasket and 16 bolts were used to secure the hemispheres together with a leak tight seal. Threaded rods were used for four of the 16 bolts to provide support legs for the sensor. A hole was drilled in the top and bottom of the sensor. The bottom hole had a liquid tight multi-hole cord grip installed to allow the cables to exit. The top hole was left open for filling and draining of the sensor. A gap of 0.25 inch between the plates was used. A data acquisition system from Tech4Imaging, LLC was used to collect data for each sensor.

The plates for the octahedron design were made from 0.75 thou thick adhesive backed copper foil. The plates were adhered to the surface of the sensor and ¼ inch gaps were cut between them. Solder connections to each plate were made near the poles, the coaxial cable ground shield was terminated 0.5 inch from the plate and the wires were staked with epoxy to prevent movement.

The dodecahedron design is more complex and requires a different plate construction method. A thicker 0.003 inch copper was selected. It was press-formed in a 3D printed mold to the curvature of the sphere and trimmed to size. The 3 plates near each pole were epoxied into place. The plates near the equator protrude past the edge of the hemisphere and therefore needed support. These plates were epoxied to the outer diameter of curved acrylic plates and then epoxied to the hemisphere, allowing all the copper plates to be on the same radius. The wires for the equatorial plates were routed in between the polar plates back to the poles of the sensor. The constructed sensor designs are shown in Fig. 4.

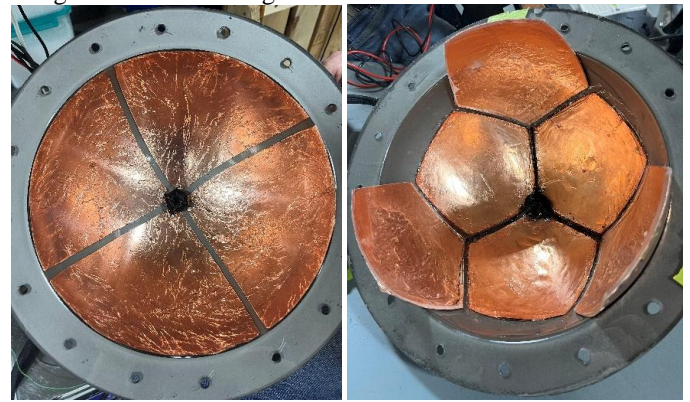

Design 1            Design 2

Fig. 4. Constructed Sensors

## IV. SIMULATION

The two designs were simulated in COMSOL Multiphysics to assess the sensitivity properties of each electrode layout design. The model is simulated within the 'Electric current interface' of COMSOL Multiphysics. A 2 MHz AC signal is applied for evaluation of electric field and capacitance data. The interface assumes quasistatic approximations internally, which simplifies Maxwell's equations into Gauss' equation $\nabla \cdot (\sigma + j\omega\varepsilon)(\nabla\varphi) = 0$ as the governing physics. Here, $\sigma$ and $\varepsilon$ denote the conductivity and permittivity of the material. For the present cryogen simulant, the conductivity is set to zero whereas the relative permittivity is set to 2.2. Mesh size is set to 'Normal' within the simulation interface and a parameter sweep is used to evaluate the mutual capacitance among all the plates. The sensitivity distribution is plotted in Fig. 5 by normalizing the sensitivity of each channel and then summing all of the channels together.

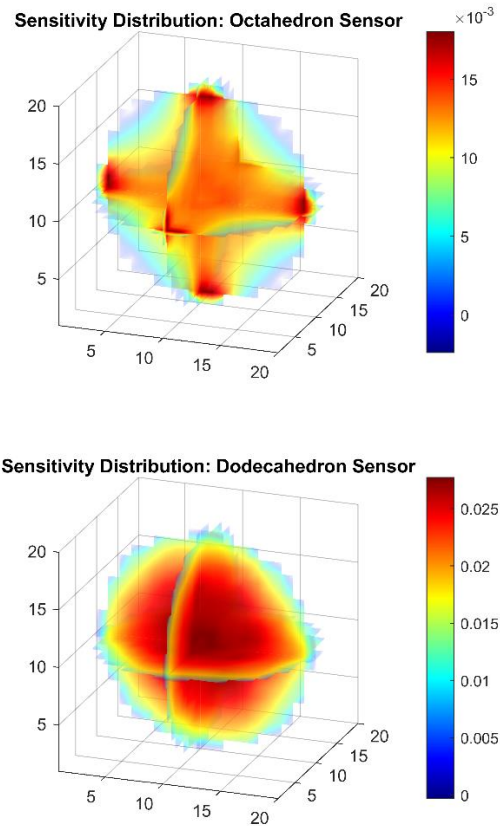

Fig. 5. Sensitivity Distributions of Sensor Designs

The sensitivity distribution plot highlights specific issues with the octahedron design, The large plates and large adjacent boundaries create large regions of negative sensitivity that are illustrated by the "dead zones" in the distribtion plot. The increased rotational symmetry of the dodecahedron design homogenizes the sensitivity distribution, minimizing the change in sensitivity as with respect to fluid position.

TABLE II
SPATIAL SENSITIVITY OF CHANNEL TYPES

| Sensor | Channel Type | Spatial Sensitivity Quotient |
|---|---|---|
| Octahedron | Adjacent | 0.0275 |
| | Semi-Adjacent | 0.0484 |
| | Opposite | 0.2246 |
| | Total Sum | 1.8096 |
| Dodecahedron | Adjacent | 0.0302 |
| | Cross | 0.0615 |
| | Opposite | 0.1662 |
| | Total Sum | 3.7479 |

Due to the spherical symmetry of the design, there are a finite number of relative plate configurations, as illustrated in Fig. 6 Also, in Fig. 6 the quantity of each type of plate configuration is listed. The Octahedron Sensor has adjacent plates that share one edge, semi-adjacent plates that share one vertex, and opposite plates that share no edges or vertices. The Dodecahedron has adjacent plates that share one edge, cross plates that share no edges or vertices, and opposite plates that are antipodes. These channel types have similar sensitivity and similar response profiles. The relationships between the channel types can provide important information about the fluid state. A high adjacent to cross channel signal ratio would indicate the fluid is in an annular configuration and a low ratio would indicate a ball of fluid in the middle. Use of these parameters in mass gauging algorithms could improve the accuracy with fluid position.

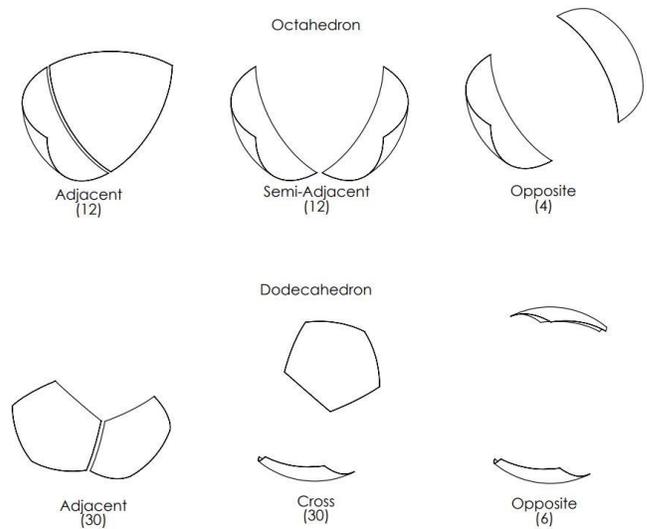

Fig. 6. Diagrams Showing Representative Channel Types

The spatial sensitivity of each channel type is calculated according to (3) and listed in Table II. The dodecahedron adjacent and cross channels have a higher spatial sensitivity than the octahedron, but the opposite channels are lower. However, when adding the spatial sensitivity together to determine the overall function of the sensor, the dodecahedron design is clearly more sensitive to fluid mass and has less variation with fluid position.

The DR of each sensor design can also be simulated by filling the RoI with a dielectric mineral oil for the full state and a gas phase for the empty state. This data is plotted in Fig. 7. It can be seen that, due to the larger plate area, the octahedron sensor has higher empty capacitance values and a higher DR for all channel types. The high adjacent empty capacitance and DR can be deceptive. It would appear at first glance that the adjacent channels perform best. However, as was illustrated in Fig. 3, the increase in SNR for adjacent channels comes at the expense of confined spatial sensitivity.

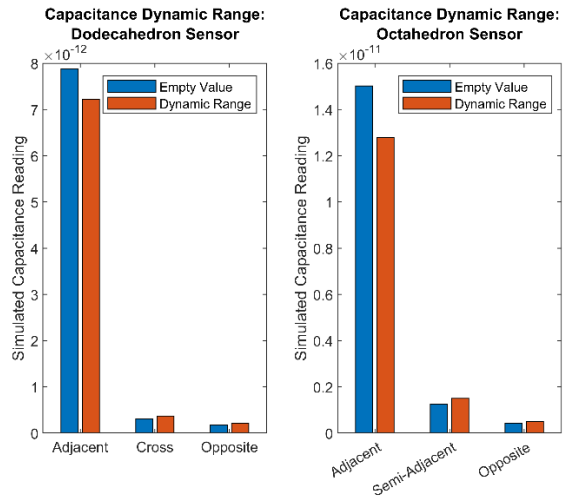

Fig. 7. Simulated Dynamic Range of Sensors by Channel Type

## V. Experiment

The primary propellants of interest for microgravity gauging are cryogenic liquid oxygen and liquid hydrogen. These materials have very low dielectric constants as shown in Table III, making them challenging to distinguish from gas. A suitable less explosive test analog is needed. Liquid nitrogen has a dielectric constant between hydrogen and oxygen, but operations at cryogenic temperatures require a more robust test fixture, which was not the focus of this study. Special heat transfer fluids like Galden HT-55 have low dielectric constants and have been used in other studies, [15] but at the time of this study were difficult to obtain due to supply issues. A standard mineral oil was ultimately selected because it is safe, readily available, and can be used at room temperature without special equipment. The disadvantage is the higher dielectric constant which will provide a higher signal than the true propellants.

TABLE III
DIELECTRIC CONSTANTS OF RELEVANT MATERIALS

| Material | Dielectric Constant | Temperature (K) |
| --- | --- | --- |
| Hydrogen ($H_2$) [28] | 1.228 | 20.4 |
| Oxygen ($O_2$) [28] | 1.507 | 80 |
| Nitrogen ($N_2$) [28] | 1.454 | 70.15 |
| Mineral Oil[1] | 2.16 | 296.15 |
| Galden HT-55 [29] | 1.86 | 298.15 |

A series of stratified filling experiments illustrated in Fig. 8 were conducted to measure the response of the sensors to volume fraction. A sensor was placed on a scale and then repeatedly filled with a discrete amount of mineral oil. At each fill level, a mass reading was recorded along with a 60 second capacitance dataset. Then, to give an indication as to the rotational stability, each sensor was tilted to 45 degrees and the filling repeated. Finally, a 3.94" diameter polypropylene ball filled with oil was advanced from the top to the bottom of the tank in ¼" increments to determine the positional stability of the design. The sensor and ball position test fixture are shown in Fig. 9. A polypropylene rod is attached to the ball and marked with graduations. A clamp is used to suspend the ball at specific heights inside the spherical tank.

[1] Measured in-house using Brookhaven BI-870 Dielectric Constant Sensor

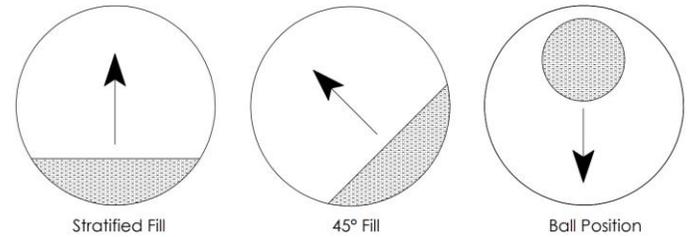

Fig. 8. Diagrams of Fluid Fill During Experiments

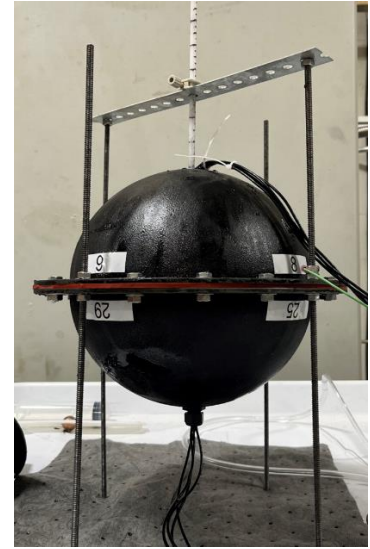

Fig. 9. Ball Position Test Fixture

The signal gain, DR, and SSNR for each measurement channel are naturally different due to small imperfections in the prototype design. Each channel is therefore normalized between the full and empty measurements to scale the datasets for comparison. The DR and SSNR are averaged together by channel to provide the DR and SSNR of the sensor.

## VI. Results and Discussion

The Octahedron stratified fill data is shown, organized by channel type, in Fig. 10. The opposite channels and semi-adjacent channels have a similar fill curve shape. The magnitude increases linearly with fill level until it approaches the equator. Then, as a result of the singularities at the equator, the magnitude experiences a large increase in signal as the fluid level passes through. As filling continues, there is almost no change in response from 0.6 to 1. The adjacent channels overall have an erratic response. Adjacent channels on the bottom of the hemisphere peak at a mass fraction around 0.5, well above the full normalization value, then decrease as filling continues. Adjacent channels in the top hemisphere have very little signal change until a mass fraction of 0.5 where they then increase dramatically. When the test is repeated at 45° in Fig. 11, similar results are observed. Adjacent plates at the bottom have sharp increases in capacitance at low mass fractions. Adjacent channels also peak as they become submerged and then decrease in magnitude as filling continues. Opposite and semi-adjacent channels have a smoother curve because the singularities are more distributed and do not contact the fluid level all at once.

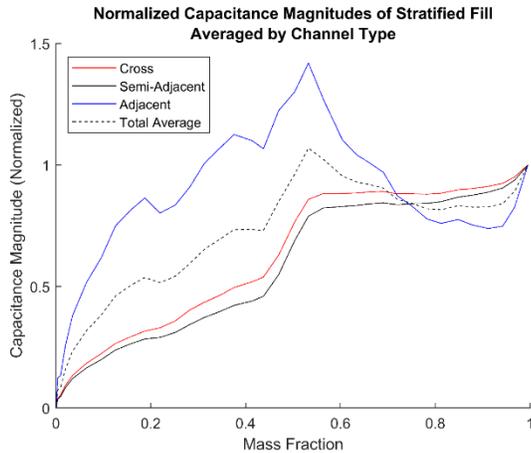

Fig. 10. Octahedron Experimental Stratified Fill Data: Averaged by Channel Type

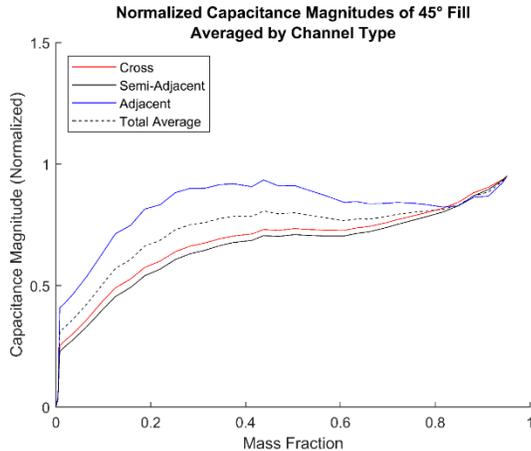

Fig. 11. Octahedron Experimental 45° Fill: Averaged by Channel Type

Next the effect of the ball position is measured, by advancing the mass ¼" at a time through the sensing region. In order to insert the ball apparatus into the sensing region, the sensor had to be disassembled and re-assembled. Small physical changes cause relatively large shifts in the normalization values, and due to the presence of the ball, the sensor cannot be normalized conventionally. For this test, the data was zeroed to the initial value and scaled based on the DR of the stratified fill test. Based on this normalization method, a y-axis value of 1 represents a change equivalent to 100% of the DR. This data is averaged by channel type in Fig. 12, demonstrating the poor overall performance, even if adjacent channels are discarded. The Opposite and Semi-Adjacent Channels have a shift of nearly the entire DR, and the adjacent channels well exceed the DR of the sensor. This is mainly due to the low DR and low SNR of the sensor, allowing small noise and position changes to dramatically affect the overall measurement.

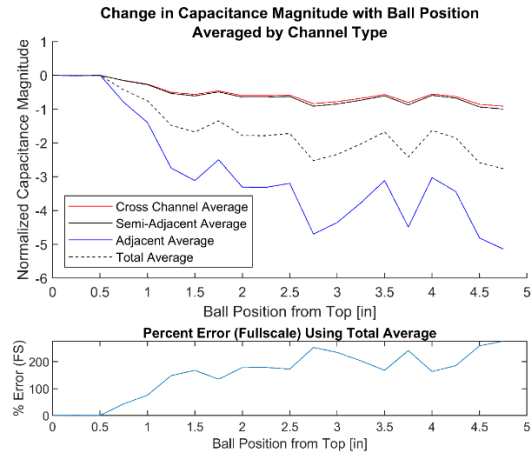

Fig. 12. Octahedron Experimental Ball Position Test: Averaged by Channel Type

The dodecahedron stratified fill data is presented in Fig. 13. Compared to the octahedron, the average dodecahedron response is highly linear. The difference in response to the 45° fill in Fig. 14 is distinguishable as the position of the singularities changes but is not nearly as substantial as the octahedron.

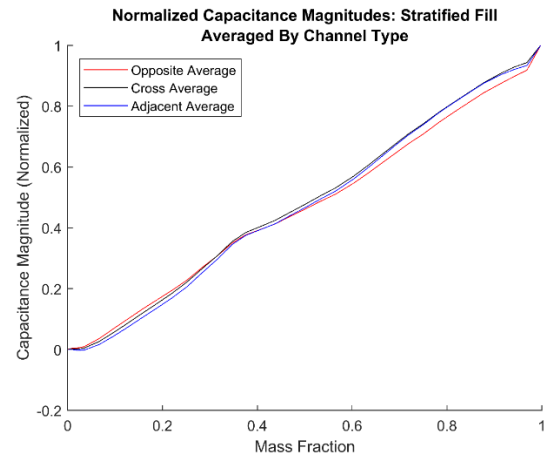

Fig. 13. Dodecahedron Experimental Stratified Fill: Averaged by Channel Type

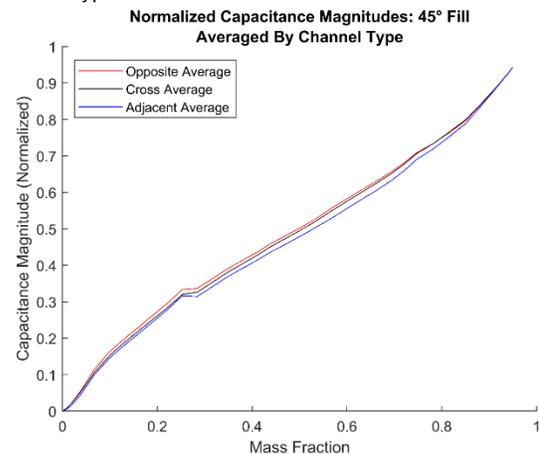

Fig. 14. Dodecahedron Experimental 45° Fill: Averaged by Channel Type

The ball position test is conducted using the same method as the octahedron sensor, with the results presented in Fig. 15. Small signal changes are observed as the ball changes position, but they have a low effect on the overall signal, with a maximum error of 0.75%.

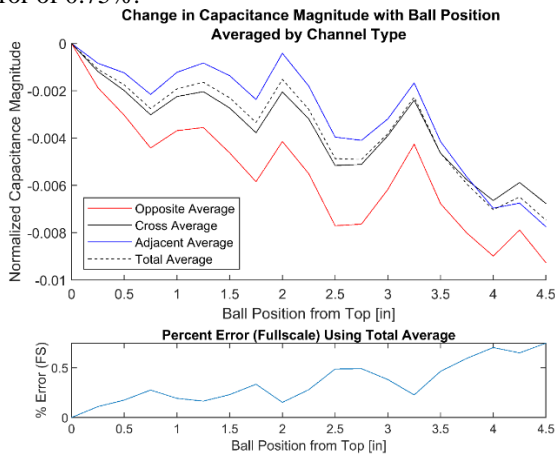

Fig. 15. Dodecahedron Experimental Ball Position Test: Averaged by Channel Type

TABLE IV
DR AND SSNR OF DESIGNS

| Sensor | Channel Type | DR | SSNR |
|---|---|---|---|
| Octahedron | Adjacent | 54.58 | 0.152 |
| | Semi-Adjacent | 192.02 | 1.005 |
| | Opposite | 223.13 | 3.232 |
| | Total Average | 137.56 | 0.958 |
| Dodecahedron | Adjacent | 2956.66 | 19.099 |
| | Cross | 3302.26 | 42.548 |
| | Opposite | 3056.31 | 98.430 |
| | Total Average | 3162.33 | 36.970 |

The reduced impact of fluid position on the measurement is a significant advantage of the dodecahedron design, but it is also accompanied by a large increase in DR and SSNR as shown in Table IV. It should be noted however, that the experimental DR and SSNR are dependent not only on the change in capacitance but the sensor construction, grounding, and electronic gain, all of which could be optimized to improve the performance of both designs. In this case, the standing capacitance of the channels in the octahedron design was high compared to the dodecahedron which required lower gain, resulting in a lower DR. The reduced impact of fluid position, however, is less dependent on experimental setup and is clearly superior in the dodecahedron design.

The capacitance data is analyzed through an averaging approach (ECVS), where normalized capacitance magnitudes are averaged together and through an imaging approach (ECVT), where a sensitivity matrix is used to inverse-solve an image of the fluid position in 3D space. The voxel density is then summed to create a volume fraction reading. The performance of each method is compared. ECVT image reconstruction is an inverse problem, where a large array of voxels, in this case 20x20x20, is reconstructed from a much smaller number of independent measurements. The Octahedron sensor has 28 independent channels, and the Dodecahedron has 66. The large discrepancy of independent measurements to reconstructed voxels results in an ill-posed reconstruction problem. A variety of reconstruction algorithms exist to address this complex inverse problem [30] and two options are considered below.

ECVS is calculated as follows. First the Adjacent channels are discarded to simplify the calculation as they are sensitive to only a small fraction of the tank volume, have a high rotational instability, and in some regions a decreasing response as the fill increases. Then the capacitance magnitudes for the stratified fill and 45° fill are fit with a 3rd order polynomial, heavily weighted at the endpoints. A percent ± error due to fluid rotation can be generated by taking the maximum difference between the two fill profiles and the polynomial fit. It should be noted that this fit covers only 2 fill profiles, and due to the uneven sensitivity inside the sphere, all extreme fill types should ideally be accounted for to provide a better estimated error. An annular fill from the outside towards the center of the sphere, a ball fill from the center towards the radius, and a variety of unique angular stratified fills would cover the most extreme cases and provide a better error calculation but are difficult experiments to perform in earth gravity.

The averaged octahedron fill profile is presented in Fig. 16. A best-fit polynomial is generated for the two fill cases with the capacitance signal as the input and the mass fraction as the output. The polynomial could also be used as a calibration curve for the sensors to relate the capacitance reading to the mass fraction. The instrument error with fluid rotation is the maximum difference at a given capacitance magnitude of indicated mass from real mass. Due to the large effect of the singularities in the octahedron design there is around ±15% error across the range of mass fractions.

The dodecahedron profile in Fig. 17 has a dramatically lower error. The more limited effect of singularities is seen as the curves diverge between 0 and 30% fill but the error decreases at higher fill levels. The maximum error for this design is around ±5%.

An explanation for the differing effects of singularities can be found in Fig. 18. During the stratified fill case, the octahedron encounters most singularities all at once, at 50% fill. The 45° fill case more smoothly distributes the effect of the singularities, but the low symmetry of the design lead to wide discrepancies in fill curve profiles. The dodecahedron on the other hand, encounters 15% of singularities at 4.5% fill and 30% of singularities at 26% fill. There is still curve divergence but due to the greater dispersion of singularities, higher symmetry, and higher DR the overall effect on the signal is greatly diminished.

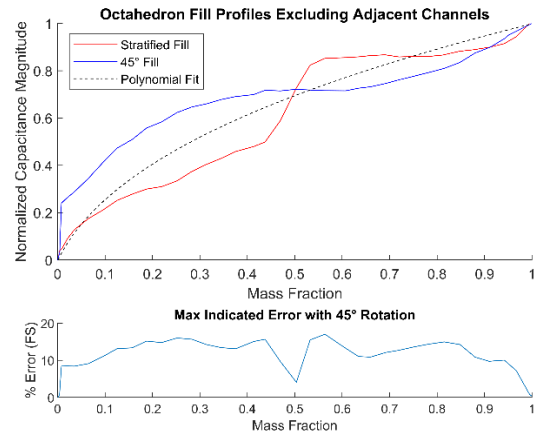

Fig. 16. Octahedron Volume Fraction Fill Profiles using ECVS Calculation Approach and Maximum Indicated Error with 45° Rotation

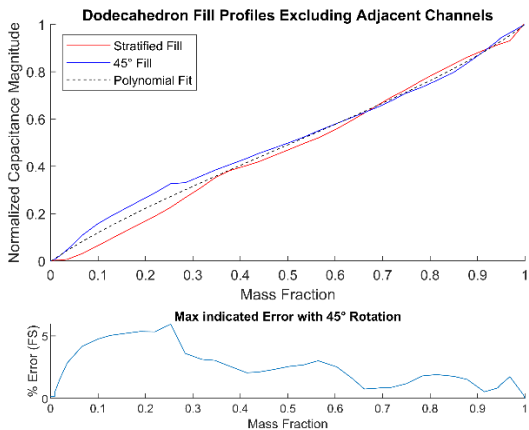

Fig. 17. Dodecahedron Volume Fraction Fill Profiles using ECVS Calculation Approach and Maximum Indicated Error with 45° Rotation

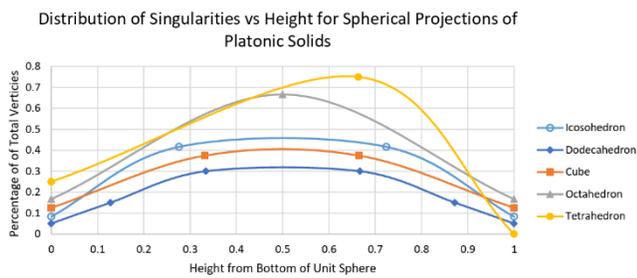

Fig. 18. Singularity Distribution for Stratified Fills of Various Spherically Tessellated Sensors

Next, the volume fraction is calculated using the ECVT imaging approach. This approach differs from the ECVS approach because it factors in a sensitivity matrix that maps the relative sensitivity of each channel to each voxel in the reconstructed image. It is more computationally intensive but has potential to be more robust to various fluid conditions. Two different algorithms are used to generate the images, Linear Back Projection and Neural Network Multicriteria Optimization Image Reconstruction Technique (NNMOIRT) [19]. Once the image is generated as a 20x20x20 voxel array, the mean voxel value is taken as the volume fraction. The NNMOIRT alpha parameter was set to 5. LBP and NNMOIRT volume fractions for the octahedron sensor are plotted in Fig. 19 and Fig. 20 and for the dodecahedron in Fig. 21 and Fig. 22. It is important to note that the error in these figures is the error due to fluid rotation only.

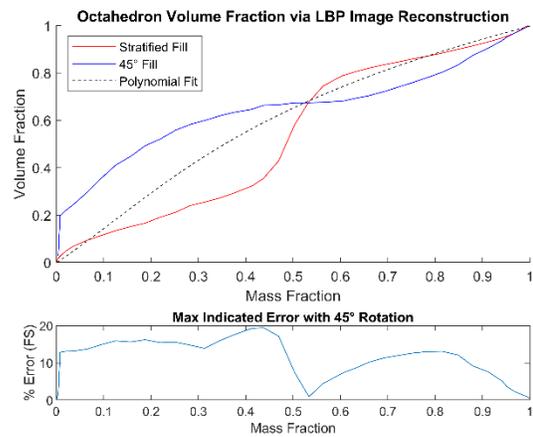

Fig. 19. Octahedron Volume Fraction Fill Profiles using LBP Image Reconstruction Calculation Approach and Maximum Indicated Error with 45° Rotation

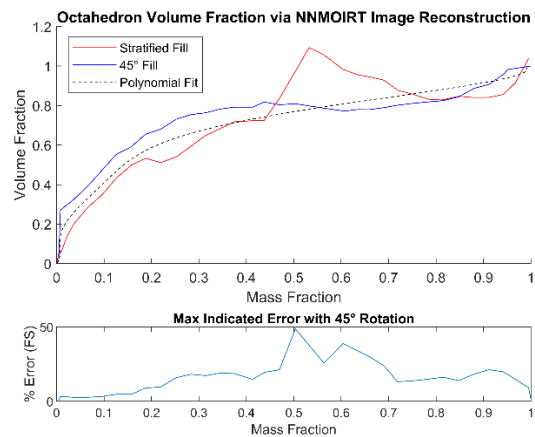

Fig. 20. Octahedron Volume Fraction Fill Profiles using NNMOIRT Image Reconstruction Calculation Approach and Maximum Indicated Error with 45° Rotation

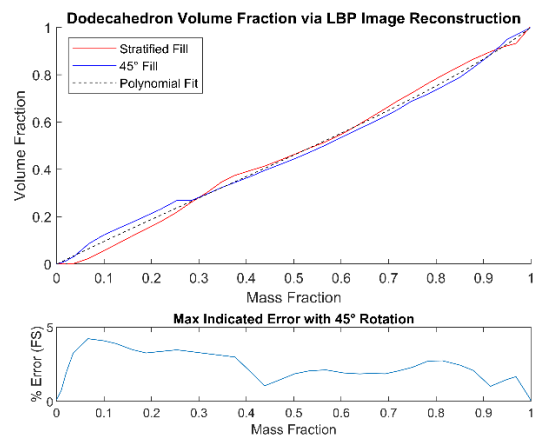

Fig. 21. Dodecahedron Volume Fraction Fill Profiles using LBP Image Reconstruction Calculation Approach and Maximum Indicated Error with 45° Rotation

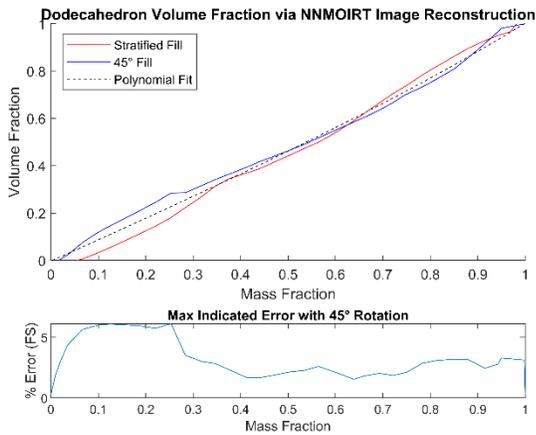

Fig. 22. Dodecahedron Volume Fraction Fill Profiles using NNMOIRT Image Reconstruction Calculation Approach and Maximum Indicated Error with 45° Rotation

The 3D image reconstructions for selected volume fractions are shown in Fig. 23 for reference. In each of these cases the stratified fill data is used to generate an image, but a stratified fluid boundary is not clearly visible in the reconstructed image. Because of the limited number of electrodes in the octahedron sensor clear images are not expected, and this is certainly indicated with the results. For both LBP and NNMOIRT reconstructions the octahedron sensor image reconstructions bear no resemblance to the stratified fluid state. The dodecahedron performs significantly better, showing a clear gradient from top to bottom. Further increasing the number of plates above twelve would increase the accuracy of the reconstructed images but would not necessarily improve the volume fraction calculation as creating a clear image of the fluid configuration and accurately measuring the mass are two different considerations that would require differently optimized sensors and reconstruction methods. Increasing the number of plates would improve the image, but could decrease the overall SSNR, especially for large tanks.

Table VI lists the accuracy of the different volume fraction calculation methods. For all cases the dodecahedron has dramatically improved accuracy values over the octahedron. The LBP and averaging mass estimations are similar, with NNMOIRT having a higher max error.

The improvement in gauging performance of the dodecahedron design is not attributed only to an increased number of electrodes, but to the increased rotational symmetry of the design. Gut [16] tested similar mass gauging systems with increased number of electrodes but was unable to reduce the maximum error with tank rotation below 8% of fullscale.

Temperature was not accounted for in this investigation due to the stability of the temperature in the laboratory. However, in space applications the effect of large temperature changes will affect the dielectric constant of the fluid, and this will have to be compensated for in the instrument.

TABLE VI
ACCURACY OF VOLUME FRACTION CALCULATION METHODS

|              | Error | Averaging | LBP    | NNMOIRT |
|--------------|-------|-----------|--------|---------|
| Octahedron   | Max   | 17.02%    | 19.44% | 48.66%  |
|              | Mean  | 8.37%     | 7.90%  | 10.11%  |
| Dodecahedron | Max   | 5.94%     | 4.22%  | 6.14%   |
|              | Mean  | 1.96%     | 1.59%  | 2.40%   |

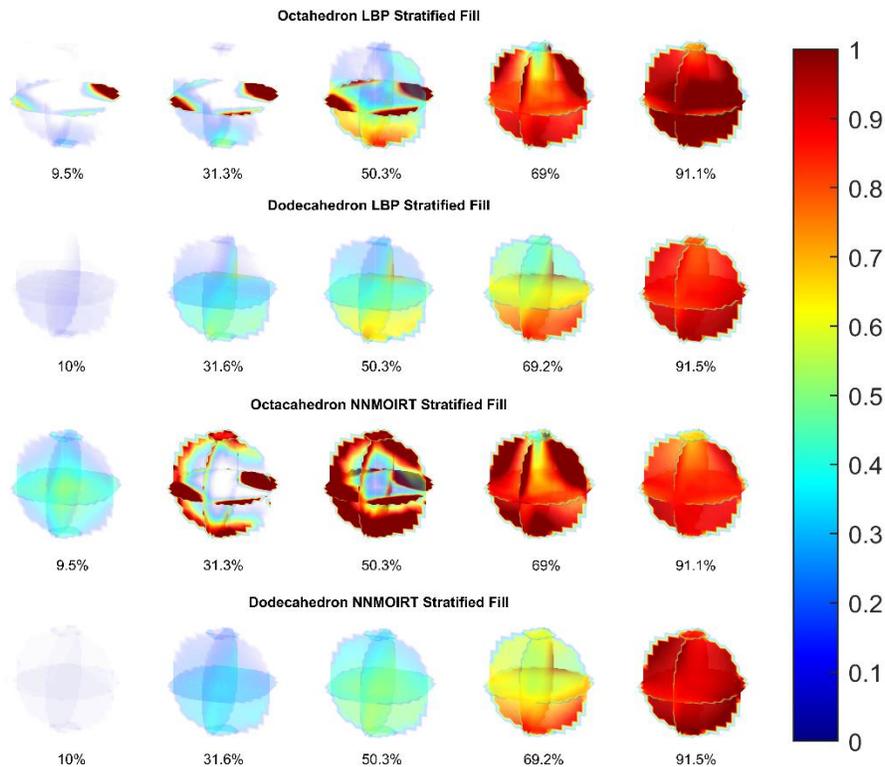

Fig. 23. Image Reconstructions for Selected Volume Fractions

## VII. Conclusion

A well-performing ECV mass gauge must have a high dynamic range, high sensitivity distribution and uniformity, a high SSNR, and high stability when a mass of fluid is moving inside the region. In the various tests discussed here, the dodecahedron sensor performed significantly better than the octahedron design. With around 5% error when the fluid is rotated 45° compared to the 15 - 20% error of the octahedron, the dodecahedron's capacitance signal is more stable with fluid position. The improvements are primarily attributed to the increase in the number of non-adjacent plates, the higher rotational symmetry, and higher overall SSNR. The DR of the octahedron sensor was lower than expected, but this may be attributable to aspects of the electronics and the large degree of curvature of the plates and the sharpness of the corners, concentrating sensitivity near the edges. The imaging resolution is better for the dodecahedron but is comparable in accuracy to the non-imaging approaches. The imaging approaches have very little advantage over averaging for the cases studied here. Imaging is intended more for fluid distribution measurement and is not optimized for mass gauging. It is possible that in microgravity, the surface tension dominated annular fluid configurations would be more accurately measured using an imaging approach, but a well-trained machine learning model would be able to compensate for fluid position as well, without introducing errors through the ill-posed image reconstruction process. Additionally, the less accurate mass gauging outcome from image reconstruction using the NNMOIRT approach is a result of using iterative techniques. Iterative techniques highlight features in the image to emphasize the fluid distribution but are not ideal for mass gauging. Whereas mass gauging is based on averaging the fluid volume fraction, iterative reconstruction is based on spatial differentiation. Nevertheless, an accurate spatial distribution obtained by iterative image reconstruction can be normalized based on EVS mass gauging. The result would be an accurate mass fraction distribution.

The dodecahedron design functions well for settled fluid configurations. It is accurate enough in the current state that it would be a beneficial addition to traditional gauging methods that lose accuracy as fill level decreases. The use of optimized machine learning or ECVT reconstruction techniques for microgravity fill cases have the potential to further improve the accuracy of the sensor.

## VIII. References

[1] F. T. Dodge, "Propellant Mass Gauging: Database of Vehicle Applications and Research and Development Studies," NASA, 2008.

[2] R. Balasubramaniam, E. Ramé and B. J. Motil, "Microgravity Liquid-Gas Two-Phase Flow: Review of Pressure Drop and Heat Transfer Correlations and Guidelines for Equipment Operability," National Aeronautics and Space Administration, Cleveland, 2019.

[3] B. Yendler, "Review of Propellant Gauging Methods," in *44th AIAA Aerospace Sciences Meeting and Exhibit*, Reno, Nevada, 2006.

[4] G. A. Zimmerli, K. R. Vaden, M. D. Herlacher, D. A. Buchanan and N. T. Van Dresar, "Radio Frequency Mass Gauging of Propellants," NASA/TM-2007-214907, Cleveland, 2007.

[5] K. M. Crosby, R. J. Werlink and E. A. Hurlbert, "Liquid Propellant Mass Measurement in Microgravity," *Gravitational and Space Research*, vol. 9, no. 1, pp. 50-61, 2021.

[6] National Aeronautics and Space Administration, "2020 NASA Technology Taxonomy," National Aeronautics and Space Administration, 2020.

[7] A. Simonini, M. Dreyer, A. Urbano, F. Sanfedino, T. Himeno, P. Behruzi, M. Avila, J. Pinho, L. Peveroni and J.-B. Gouriet, "Cryogenic propellant management in space: open challenges and perspectives," *npj microgravity*, vol. 10, no. 1, p. 34, 2024.

[8] G. A. Zimmerli, "Propellant Gauging for Exploration," in *54th JANNAF*, Denver, CO, 2017.

[9] W. Yang, "Design of Electrical Capacitance Tomography Sensors," *Measurement Science and Technology*, vol. 21, no. 4, p. 042001, 2010.

[10] A. Wang, "Electrical Capacitance Volume Tomography and its Applications in Multiphase Flow Systems," PhD Dissertation, The Ohio State University, Columbus, OH, 2015.

[11] C. Biagi, M. Nurge, R. Youngquist, J. M. Storey, L. Bird, A. Atkins, A. Swanger, N. Spangler, M. Mercado, J. Defiebre, A. Esparza and J. Hartwig, *Capacitive Mass Gauging for Fluids in Micro-g,* Kailua-Kona, Hawaii: 30th Space Cryogenics Workshop, 2023.

[12] W. Q. Yang, D. M. Spink, J. C. Gamio and M. S. Beck, "Sensitivity Distributions of Capacitance Tomography Sensors with Parallel Field Excitation," *Measurement Science and Technology*, vol. 8, no. 5, pp. 562-569, 1997.

[13] S. M. Chowdhury, M. A. Charleston, Q. M. Marashdeh and F. L. Teixeira, "Propellant Mass Gauging in a Spherical Tank under Micro-Gravity Conditions Using Capacitance Plate Arrays and Machine Learning," *Sensors*, vol. 23, no. 20, p. 8516, 2023.

[14] R. Drury, "Model guided capacitance tomography a Bayesian approach to flow regime independent

multiphase flow measurement," Doctoral Thesis, Coventry University, 2019.

[15] S. H. Yang, Y. S. Kim, N. G. Dagalakis and Y. Wang, "Flexible Assemblies of Electrocapacitive Volume Tomographic Sensors for Gauging Fuel of Spacecraft," *Journal of Spacecraft and Rockets,* vol. 58, no. 2, pp. 499-504, 2021.

[16] Z. Gut, "Using Electrical Capcitance Tomography System for Determination of Liquids in Rocket and Satellite Tanks," *Transactions on Aerospace Research,* vol. 2020, no. 1, pp. 18-33, 2020.

[17] A. Wang, Q. Marashdeh and L.-S. Fan, "ECVT Imaging and Model Analysis of the Liquid Distribution inside a Horizontally Installed Passive Cyclonic Gas-Liquid Separator," *Chemical Engineering Science,* vol. 141, pp. 231-239, 2016.

[18] P. Behruzi, A. Hunt and R. Foster-Turner, "Evaluation of Liquid Sloshing using Electrical Capacitance Tomography," in *AIAA Propulsion and Energy Forum*, VIRTUAL EVENT, 2020.

[19] W. Warsito, Q. Marashdeh and L.-S. Fan, "Electrical Capacitance Volume Tomography," *IEEE Sensors Journal,* vol. 7, no. 4, pp. 525-535, 2007.

[20] M. Atiyah and P. Sutcliffe, "Polyhedra in Physics, Chemistry and Geometry," *Milan Journal of Mathematics,* vol. 71, pp. 33-58, 2003.

[21] J. M. Storey, B. S. Marsell, M. T. Elmore and S. Clark, "Propellant Mass Gauging in Microgravity with Electrical Capacitance Tomography," National Aeronautics and Space Administration, Kennedy Space Center, Florida, 2023.

[22] F. Wang, Q. Marashdeh, L.-S. Fan and W. Warsito, "Electrical Capacitance Volume Tomography: Design and Applications," *Sensors,* vol. 10, no. 3, pp. 1890-1917, 2010.

[23] W. Q. Yang and L. Peng, "Image Reconstruction Algorithms for Electrical Capacitance Tomography," *Measurement Science and Technology,* vol. 14, no. 1, p. R1, dec 2002.

[24] Q. Zhao, J. Li, S. Liu, G. Liu and J. Liu, "The Sensitivity Optimization Guided Imaging Method for Electrical Capacitance Tomography," *IEEE Transactions on Instrumentation and Measurement,* vol. 70, pp. 1-15, 2021.

[25] G. Villares, L. Begon-Lours, C. Margo, Y. Oussar, J. Lucas and S. Hole, "A Non-linear Model of Sensitivity Matrix for Electrical Capacitance Tomography," *2012 Annual Meeting of the Electrostatics Society of America,* 2012.

[26] J. Ye, H. Wang and W. Yang, "Characterization of Electrical Capacitance Tomography Sensors with Different Diameter," *IEEE Sensors,* vol. 14, no. 7, pp. 2240-2251, 2014.

[27] W. Q. Yang and W. F. Conway, "Measurement of Sensitivity Distributions of Capacitance Tomography Sensors," *Review of Scientific Instruments,* vol. 69, no. 1, pp. 223-236, 1998.

[28] D. R. Lide, CRC Handbook of Chemistry and Physics 73rd Edition, Boca Raton, Florida: CRC Press, Inc., 1992-1993.

[29] Solvay Specialty Polymers, *Galden HT PFPE Heat Transfer Fluids,* 2013.

[30] R. K. Rasel, S. M. Chowdhury, Q. M. Marashdeh and F. L. Teixeira, "Review of Selected Advances in Electrical Capacitance Volume Tomography for Multiphase Flow Monitoring," *Energies,* vol. 15, p. 5285, 2022.

[31] S. M. Chowdhury, Q. M. Marashdeh, F. L. Teixeira and L.-S. Fan, "Electrical Capacitance Tomography," in *Industrial Tomography 2nd Edition*, Cambridge, Elsevier, 2022, pp. 3-29.

[32] A. Wang, Q. M. Marashdeh, F. L. Teixeira and L.-S. Fan, "Electrical Capacitance Volume Tomography: A Comparison between 12 and 24 Channels Sensor Systems," *Progress in Electromagnetics Research M,* vol. 41, pp. 73-84, 2015.

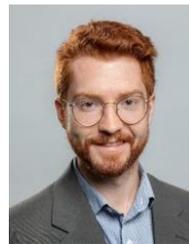

**Matthew A. Charleston** received the B.S. degree in Mechanical Engineering from The Ohio State University, Columbus, Ohio, USA in 2019. He is currently a Senior Product Development Engineer for Tech4Imaging, LLC, in Columbus, Ohio researching electrical tomography applications for multiphase steam, cryogenic, and oil flows in addition to capacitance mass gauging. He has 5 years of experience in flow meter development, previously working on advancing thermal mass flow meters and developing thermal property concentration sensors. His research interests include sensor design and optimization, multiphase flow systems, and electrical tomography.

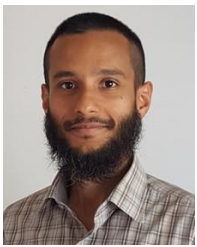
**Shah M. Chowdhury** received the B.S. degree from Bangladesh University of Engineering and Technology, Dhaka, Bangladesh, in 2012 and the Ph.D. degree from The Ohio State University, Columbus, OH, USA, in 2021, all in electrical engineering. From 2012 to 2015, he was with Samsung R\&D Institute Bangladesh as an Embedded Software Developer. He is currently a Post Doctoral Researcher at Tech4Imaging, LLC in Columbus, Ohio. His research interests include electrical tomography methods, industrial flow velocity profiling, and inverse problems.

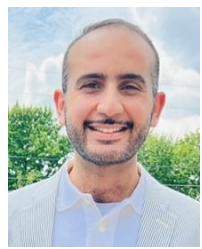
**Qussai M. Marashdeh** received the B.S. degree in electrical engineering from the University of Jordan, Amman, Jordan, in 2001, and both the M.S. and Ph.D. degrees in electrical engineering while affiliated with the Electroscience Laboratory, The Ohio State University, Columbus, in 2003 and 2006, respectively. He also received the M.S in chemical engineering and the M.B.A. from The Ohio State University in 2009 and 2012, respectively. He is cofounder, President, and CEO of Tech4Imaging LLC, in Columbus, Ohio, a startup company aimed at advancing capacitance sensing and tomography technology and their applications. His research interests include electrical tomography systems, mass flow instrumentation, electrostatics, optimization, multiphase flow, and inverse problems.

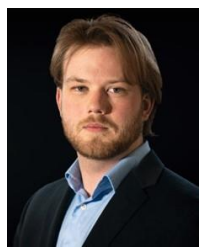
**Banjamin J. Straiton** received the B.S. degree in Electrical and Computer Engineering from The Ohio State University, Columbus, Ohio, USA in 2016. He is currently a product development manager for Tech4Imaging, LLC, Columbus, Ohio advancing electrical tomography systems for industrial multiphase flow applications. His research interests include electrical tomography and multiphase flow systems.

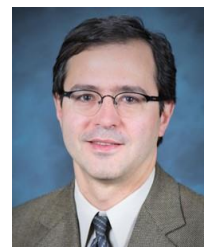
**Fernando L. Teixeira** received the Ph.D. degree from the University of Illinois at Urbana–Champaign, Champaign, IL, USA, in 1999. He was a Postdoctoral Associate with the Massachusetts Institute of Technology, Cambridge, MA, USA, from 1999 to 2000. He joined the Ohio State University (OSU), Columbus, OH, USA, in 2000, where he is currently a Professor with the Department of Electrical and Computer Engineering and is also affiliated with the ElectroScience Laboratory. His current research interests include applied electromagnetics and sensor physics. His contributions have been recognized by several awards, including the NSF CAREER Award, the triennial Booker Fellowship from the U.S. Committee of the International Union of Radio Science, the Outstanding Young Engineer Award from the IEEE Microwave Society (MTT-S), and NASA Certificates of Appreciation. He served as an Associate Editor for IEEE ANTENNAS AND WIRELESS PROPAGATION LETTERS and IET Microwaves, Antennas, and Propagation, and as a Guest Editor for Remote Sensing, IEEE JOURNAL ON MULTISCALE AND MULTIPHYSICS COMPUTATIONAL TECHNIQUES, and Progress in Electromagnetics Research.